\documentclass[ floatfix]{revtex4}
\usepackage{amsmath}
\usepackage{color}
\usepackage{amssymb}

\usepackage{graphicx}

\begin{document}

\title{ Upper critical field in  ferromagnetic metals with triplet pairing }

\author{V.P.Mineev$^{1,2}$}
\affiliation{$^1$Univ. Grenoble Alpes, CEA, INAC, PHELIQS, GT, F-38000 Grenoble, France\\
$^2$Landau Institute for Theoretical Physics, 142432 Chernogolovka, Russia}

\begin{abstract}
The theory of triplet superconductivity in ferromagnetic metals based on electron-electron interaction by spin fluctuation exchange is developed.
The equations for the upper critical field temperature dependence are derived. In contrast  to the similar  equations for the superconductivity in 
two band metals  they contain the pairing amplitudes and the  Fermi velocities   depending on magnetic field. The critical field behaviour  near the critical temperature and at $T=0$ is established analytically.
 \end{abstract}

\date{\today}
\maketitle

\section{Introduction}

In the conventional superconductors the pairing interaction arises from the coupling between electrons and the lattice.
The theory of superconductivity based on electron-phonon interaction formulated first by G.M.Eliashberg \cite{Eliashberg1960} more than half century ago has been transformed now in well developed part of condensed matter physics \cite{Marsiglio2008}.
The superconductivity in ferromagnetic uranium compounds UGe$_2$, URhGe and UCoGe  apparently does not belong to conventional type
(see the recent  review by D.Aoki, K.Ishida and J.Flouquet \cite{Aoki2019}). 
  The superconducting states in these materials usually are developed at temperatures much lower than the Curie temperature  and characterised by extremely high upper critical fields what points out that here we are dealing with the Cooper pairing with parallel electron spins.
   All three uranium compounds have orthorhombic crystal structure with an inversion center.
Due to the Pauli principle the spin-triplet superconducting order parameter  in a centre-symmetric medium should be odd function of pairing electrons momenta. The simple estimation shows that the amplitude of triplet component of electron-phonon pairing interaction is by a factor of $\sim (T_{sc}/\Theta_D)^2 $  smaller than the  singlet pairing amplitude.
Here $T_{sc}$  is the critical temperature of superconducting transition and $\Theta_D$ is the Debye temperature.
This leaves no hope for the electron-phonon pairing mechanism of 
odd parity triplet superconductivity.

There are several theoretical approaches to the description of superconductivity mechanism  in ferromagnetic metals.
Some theoretical studies have used pairing interaction induced by the spin polarisation in
 itinerant ferromagnetic  Fermi liquid   similar to the interaction by paramagnon exchange widely accepted in the theory of superfluid $^3$He \cite{Nakajima1973}. First it was done for the isotropic electron liquid  in the weak coupling static limit \cite{Fay1980}.
 Dynamics of spin polarisation  and the system anisotropy have been taken into consideration in semi-phenomenological treatments of pairing interaction 
in  itinerant ferromagnetic  Fermi liquid  in the papers \cite{Monthoux2001,Tada2011,Tada2013}. 
In the other approaches there were used 
a pairing  interaction  owing to magnetic excitations exchange  between conducting electrons and
 an artificially introduced  ferromagnetic subsystem formed by localised magnetic moments,  first  in strong  external field perpendicular to spontaneous magnetisation \cite{ Hattori2013}, then in the absence of external field  and in neglect of  internal magnetic field interaction with electron charges \cite{Bulaevskii2019}.
  
  The  pairing mechanism determined  only by transverse dynamics of magnetic degrees of freedom studied in  \cite{Bulaevskii2019} leads to the
 of absence of pairing in
the limit of zero temperature due to freezing out  magnetic excitations with energy gaped by the anisotropy. 
This statement is in fact  too strong. One can demonstrate that the static limit of frequency dependent equations for the spin-up and the spin-down component of the order parameter   obtained in \cite{Bulaevskii2019}  exactly corresponds  to  the equations induced by the static transverse susceptibility components derived in \cite{Mineev2017}. The latter have the finite solution for the critical temperature at any constant of pairing interaction. However,  due to the smallness of magnetic susceptibility in the directions perpendicular to the easy axis the critical temperature is really exponentially low.

Besides the transverse magnetisation  dynamics one must  take into account
the longitudinal fluctuations of magnetisation\cite{Karchev2003}.
 The corresponding frequency dependence of longitudinal susceptibility is \cite{Mineev2017,Mineev2013} 
\begin{equation}
\chi_{zz}({\bf k},\omega)=\frac{A}{-i\omega+\Gamma_{\bf k}},~~~~~~~~~
\Gamma_{\bf k}=2A\left[-2\alpha_z+\gamma^z_{ij}k_ik_j\right],~~~~\alpha_z=\alpha_0(T-T_{Curie}).
\label{chi}
\end{equation}
According to the experiments \cite{Huxley2003,Stock2011} the magnitude of $\Gamma_{{\bf k}\to 0}$ at temperatures substantially lower than $T_{Curie}$  is of the order of several Kelvin. So, in the temperature region where the superconducting state is realised one can neglect  the frequency dependence of longitudinal susceptibility. In the UCoGe, as well in the other uranium compounds,   the static susceptibilities  in the transverse directions $\chi_{xx}(\omega=0),~\chi_{yy}(\omega=0)$ are much smaller than the longitudinal static susceptibility $\chi_{zz}(\omega=0)$ \cite{Huy2008,Hardy2011} . The latter according to the papers
\cite{Mineev2011,Mineev2017} serves as the main source of pairing. 
 The theory of triplet pairing 
 developed in the papers \cite{Mineev2011, Mineev2017} 
qualitatively explains the interplay between the pressure dependence of the Curie temperature and the critical temperature of the superconducting transition and several other
observations,   in particular, the peculiar upward curvature of the upper critical field parallel to spontaneous magnetisation in UCoGe.

 Another important property of uranium ferromagnetic superconductors  is that the critical temperature in these materials  correlates with electron effective mass changes caused by pressure or external magnetic field \cite{Aoki2019}.
In the theory of strong coupling superconductivity the amplitude of electron-phonon pairing interaction and the effective mass renormalisation are determined by the same parameter $\lambda$ \cite{Scalapino1969,McMillan1968}.   The knowledge of this parameter allows to find 
the upper critical field temperature dependence \cite{Schlossmann1986,Bulaevskii1988,Thomas1996,Clemot1999}. If $\lambda$ itself is  magnetic filed dependent one can try to restore its value from the experimentally measured $H_{c2}(T)$. 
This has been  done  in  \cite{Beilun2017} in the assumption that  the upper critical field temperature dependence is determined by the expression identical to the obtained in the electron-phonon  interaction theory. For the magnetic field parallel to spontaneous magnetisation $\lambda(H)$ dependence  in UCoGe found by this procedure by Beilun Wu and co-authors 
 \cite{Beilun2017} occurs in reasonable correspondence with magnetic field dependence of the specific heat 
$\left [C(T)/T\right]\sim [1+\lambda(H)]$. Moreover, the authors of  \cite{Beilun2017} have demonstrated the  correspondence of this quantity and the magnetic field dependence of the constant of pairing interaction derived in the weak coupling theory  \cite{Mineev2011, Mineev2017}. The trouble, however, is that  in a theory of electron-phonon interaction parameter $\lambda$ is field independent. On the other hand,
the static weak coupling approach  says nothing about the field dependence of electron effective mass observed experimentally \cite{Aoki2019,Beilun2017}.

Obviously, the mass renormalisation and the constant of pairing interaction must be derived in frame of the same theoretical approach. The corresponding theory of two band ferromagnetic
superconductor with triplet pairing
is developed in the present paper.
The field dependence of effective mass and the pairing amplitudes  are established in the next two Sections. In the fourth Section the equations for the upper critical field temperature dependence are derived. 
They are similar  to the  corresponding equations for an usual two band superconductor but differ from them by the field dependence of the Fermi velocities and the pairing interaction.
The critical field behaviour  near the critical temperature and at $T=0$ is established analytically.
In the conclusion I 
list the  main results as well the principal simplifications made in the process of its derivation.

\section{Electron effective mass field dependence}

Effective mass  $m$ of an electron in metal differs from the bare electron mass due to static and dynamic (electron-phonon) interaction with crystal lattice. Here we will be interested in
the additional contribution to electron effective mass arising 
due to electron-electron interaction through the spin fluctuations exchange.  It can be calculated  as 
\begin{equation}
\frac{m^{*}_a}
{m}-1=-\frac{\partial~\textrm{Re}\Sigma_a{(\bf p},\varepsilon)}{\partial\varepsilon}\left |_{\varepsilon=0} \right.,
\end{equation}
where  $\Sigma_a({\bf p},\varepsilon)$ is electron self-energy function. In lowest order in interaction with longitudinal spin fluctuations the frequency dependent one-particle self energy is
\begin{equation}
\Sigma^a{(\bf p},\varepsilon)=-i\frac{g^2}{2}\int\frac{d\omega}{2\pi}\frac{d^3k}{(2\pi)^3}G^a({\bf p}-{\bf k},\varepsilon-\omega)\chi_{zz}({\bf k},\omega),
\end{equation}
where 
 $g$ is the
coupling constant, and $\chi_{zz}(\bf k,\omega)$ is
the longitudinal  dynamical
spin susceptibility which we will treat using the phenomenological formula (\ref{chi}). In presence of external field $H$ along the easy axis the function $\Gamma_{\bf k}$  in this formula is expressed \cite{Mineev2017} through 
the field dependent magnetisation $M_z(H)$ and magnetisation in the absence of field  $M_{z0}$ (spontaneous magnetisation) as follows
\begin{equation}
\Gamma_{\bf k}=2A\left[\beta_z(6M_z^2-M_{z0}^2)+\gamma^z_{ij}k_ik_j\right].
\end{equation}
For simplicity we will work with isotropic dispersion law $\gamma^z_{ij}k_ik_j\to\gamma^z k^2$.

The Green function
of non-interacting electrons in spin-up, spin-down bands  ($a=\uparrow, \downarrow$) split by the magnetic field is 
\begin{equation}
G_{0}^{a}({\bf p},\varepsilon)=\frac{1}{\omega-\xi_a({\bf p})+
i\delta~\textrm{sgn}~
\xi_a({\bf p})},
\end{equation}
where
\begin{equation}
\xi_{\uparrow,\downarrow}({\bf p})=\varepsilon({\bf p})\mp\mu_B (h+H)-\mu
\end{equation} 
 are the electron energy counted from the chemical potential $\mu$, 
$h$ is the exchange field, $H$ is an external field parallel to it.

Making use the spectral representation
\begin{equation}
\chi_{zz}({\bf k},\omega)=-\frac{2}{\pi}\int_0^\infty\frac{ Im\chi_{zz}({\bf k},\Omega)\Omega d\Omega }{\omega^2-\Omega^2+i\delta}
\end{equation}
and performing integration over $\omega$ we obtain
\begin{equation}
\Sigma^a{(\bf p},\varepsilon)=g^2\int\frac{d\Omega}{2\pi}\int \frac{d^3k}{(2\pi)^3} Im\chi_{zz}({\bf k},\Omega)\left \{\frac{\Theta(\xi_a({\ p}-{\bf k}))}{\varepsilon -\Omega -\xi_a({\bf p}-{\bf k})+i\delta }+
\frac{\Theta(-\xi_a({\ p}-{\bf k}))}{\varepsilon +\Omega -\xi_+({\ p}-{\bf k})-i\delta }\right\}
\end{equation}
To 
find the self energy part at Fermi surface determined for quadratic electron spectrum by the equation
\begin{equation}
\frac{(p_{F}^{a})^2}{2m}\mp\mu_B (h+H)=\mu
\end{equation}
we put 
\begin{equation}
\xi_a({\bf p}-{\bf k})\approx-\frac{p_{F}^{a} k\cos\theta}{m}+\frac{k^2}{2m},
\end{equation}
where $m$ is an electron effective mass determined by the crystal lattice effects, electron-phonon interaction but not  by the spin fluctuation exchange. 
Owing to
the condition $\xi_a({\bf p}-{\bf k})>0$   the upper limit of integration over $\cos\theta$ in the first integrand term is $k/2p_{F}^{a}$. The same value serves as the lower limit in the second term. Performing the integration over solid angle we obtain
\begin{equation}
\textrm{Re}\Sigma^a{(\bf p},\varepsilon)=g^2\frac{m}{p_{F}^{a}}\int\frac{d\Omega}{2\pi}\int \frac{kdk}{(2\pi)^2} Im\chi({\bf k},\Omega)\left \{
\ln\left | \frac{\varepsilon-\Omega}{\varepsilon+\Omega}\right |+\ln\left |\frac{\varepsilon+\Omega-\frac{k^2}{2m}+\frac{kp_{F}^{a}}{m}}{\varepsilon-\Omega-\frac{k^2}{2m}-\frac{kp_{F}^{a}}{m}}
\right |\right\}
\end{equation}
Main contribution  to integral at $\varepsilon \to 0$ gives the first term $ \ln\left |\frac{\varepsilon-\Omega}{\varepsilon+\Omega}\right |\approx-\frac{\varepsilon}{2\Omega}$. Substituting this expression one can easily calculate integral over 
$\Omega$. Then, 
 performing the integration over $k$ from $0$ to some cut-off $k_c$ we obtain
\begin{equation}
\lambda_a=\frac{m^{*}_a}{m}-1=
\frac{g^2}{32\pi^2\gamma^z p_{F}^{a}}
\ln\left\{ 1+\frac{\gamma^z k_c^2}{2\beta_z(3M_z^2-M_{z0}^2)} \right\}.
\end{equation}
Thus, via  $p_{F}^{a}$ and $M_z$ field dependence the electron effective mass is the magnetic field dependent quantity.
In similar manner one can calculate the electron effective mass renormalisation due to transverse spin fluctuations. But they  have smaller magnitude due to much smaller value of susceptibilities in direction perpendicular to the easy magnetisation axis.
The  low temperature  specific heat of electron gas in two band ferromagnet is  also decreased with magnetisations growth along the easy axis
\begin{equation}
\frac{C(H)-C(H=0)}{C(H=0)}=\lambda_\uparrow+\lambda_\downarrow\approx \frac{g^2}{16\pi^2\gamma^z \sqrt{2m\mu}}
\ln\left\{ 1+\frac{\gamma^z k_c^2}{2\beta_z(3M_z^2-M_{z0}^2)} \right\}.
\end{equation}

Near the Fermi surface the Green functions  of interacting electrons are
\begin{equation}
G^{a}({\bf p},\omega)
=\frac{1}{\omega-\xi_a({\bf p}) -\Sigma_a+
i\delta~\textrm {sgn}~\omega}=\frac{(1+\lambda_a)^{-1}}{\omega-\xi^*_a({\bf p})+
i \gamma},
\label{Green}
\end{equation}
where $\xi^*_a({\bf p})=(p-p_F^a)v_F^a$ and  
\begin{equation}
v_F^a=\frac{p_F^a}{m^*_a} 
\end{equation} 
is the Fermi velocity. 
It depends on magnetic field  through the Fermi momentum and through the electrons effective mass. 

The  logarithm dependence of the electrons effective mass from magnetisation originates from our assumptions about isotropy of the electron spectrum and the spin excitations spectrum  made to perform
of all calculations analytically. In case of realistic spectra with orthorhombic anisotropy the conclusion about suppression of  electrons effective mass by the external magnetic field directed
along spontaneous magnetisation is still valid.

\section{Critical temperature}

We will consider so called equal-spin pairing state that is taking in consideration just spin up-up $\Delta^\uparrow$ and spin down-down $\Delta^\downarrow$ components of the order parameter and ignoring zero spin projection component. If pairing occurs due to spin fluctuation exchange all the pairing amplitudes expressed through the  corresponding components of 
susceptibility.
As, this was already mentioned, in the temperature region where the superconducting state is realised one can neglect the frequency dependence of susceptibilities. This case 
the linear in respect of the order parameter component equations for determination of critical temperature (upper critical field) are \cite{Mineev2017}
\begin{eqnarray}
&\Delta^{\uparrow}({\bf p},{\bf q})
=-T
\sum_{n}
\sum_{{\bf p}' }
\left\{
V^{\uparrow\uparrow}({\bf p},{\bf p}')G^\uparrow G^\uparrow\Delta^{\uparrow}({\bf p}',{\bf q})+
V^{\uparrow\downarrow}({\bf p},{\bf p}')G^\downarrow G^\downarrow\Delta^{\downarrow}({\bf p}',{\bf q})
\right\},
\label{up1}\\
&\Delta^{\downarrow}({\bf p},{\bf q})
=-T\sum_{n}\sum_{{\bf p}' }
\left\{
V^{\downarrow\uparrow}({\bf p},{\bf p}')G^\uparrow G^\uparrow \Delta^{\uparrow}({\bf p}',{\bf q})+
V^{\downarrow\downarrow}({\bf p},{\bf p}')G^\downarrow G^\downarrow \Delta^{\downarrow}({\bf p}',{\bf q})
\right\},
\label{down1}
\end{eqnarray}
where  the Green functions products have the following arguments 
$$
G^a G^a=G^a({\bf p}',\omega_n)
G^a(-{\bf p}'+{\bf q},-\omega_n),
$$
but
unlike to the paper \cite{Mineev2017} here we should take into account the mass renormalisation due to spin fluctuations and use the Green functions 
in the form given by Eq.(\ref{Green})
in the Matsubara representation.

The pairing amplitudes  are expressed through the odd part of spin  susceptibilities \cite{Mineev2017}.
In  the case of absence of longitudinal spin fluctuations studied by Bulaevskii et al \cite{Bulaevskii2019} the amplitudes $V^{\uparrow\uparrow}=V^{\downarrow\downarrow}=0$ and the order parameter components  $\Delta^{\uparrow}$,   $\Delta^{\downarrow}$ cannot exist without each other. The pairing interaction is supported  by the pairing amplitudes $V^{\uparrow\downarrow}=
2ig^2\chi^u_{xy}$, $V^{\downarrow\uparrow}
=
-2ig^2\chi^u_{xy}$  not vanishing even in the case of tetragonal symmetry assumed in the paper   \cite{Bulaevskii2019}.

We will work with equal spin pairing B-state \cite{Mineev2017} with the order parameter
\begin{eqnarray}
&\Delta_B^\uparrow({\bf p},{\bf q})=\hat p_z\eta^\uparrow({\bf q}),
\nonumber\\
&\Delta_B^\downarrow({\bf p},{\bf q})=\hat p_z\eta^\downarrow({\bf q}).
\label{B}
\end{eqnarray}
Here, and what follows, $\hat p_z$ is the $z$-component of the unit momentum vector $\hat{\bf p}={\bf p}/|{\bf p}|$.
The treatment of equal spin pairing A-state 
is much more cumbersome because its spin-up and spin-down pairing amplitudes present the linear combinations of $k_x$ and $k_y$
components and the system of equations for the critical temperature or the upper critical field determination inevitably consists of 4 equations.
The Ginzburg-Landau theory for the state (\ref{B}) is developed in the authors paper  \cite{Mineev2018}.

The  internal field acting on the electron charges in uranium ferromagnet  is small
in comparison with the upper critical field at low enough temperatures. Hence, the formal determination of critical temperature of transition to the superconducting state in zero external field can be obtained by  ignoring the coordinate (or ${\bf q}$) dependence in
 the  Eqs.(\ref{up1}),(\ref{down1}). Then the linear equations for the order parameter Eq.(\ref{B}) components are
\begin{eqnarray}
\eta^\uparrow=\left ({\cal G}_{1}^\uparrow\eta^\uparrow+
{\cal G}_{2}^\downarrow
\eta^\downarrow\right )S(T),
\nonumber\\
\eta^\downarrow=\left ({\cal G}_{2}^\uparrow\eta^\uparrow+{\cal G}_{1}^\downarrow\eta^\downarrow
\right )S(T).
\label{Ax}
\end{eqnarray}
 For the B-state  $\chi^u_{xy}$ component of susceptibility plays no role, and  
\begin{eqnarray}
{\cal G}_{1}^a=\frac{N_0^a g^2}{1+\lambda_a}\langle\chi_{zz}^{ua}(p_z,p_z)\rangle=\frac{N_0^a g^2}{1+\lambda_a}\frac {\gamma_{zz}^{z}(p_F^a)^2\langle (\hat p_z)^2\rangle}{4\left [\beta_z(3M_z^2-M_{z0}^2)+\gamma^z (p_F^a)^2\right ]^2},~~~~~~~~~~~~~~~~~~~~~~~
\label{G1}\\
{\cal G}_{2}^a=\frac{N_0^a g^2}{{1+\lambda_a}}\langle\chi^{ua}_{xx}(p_z,p_z)-\chi^{ua}_{yy}(p_z,p_z)\rangle=\frac{N_0^ag^2}{{1+\lambda_a}}
\left (\frac{\gamma_{zz}^{z}(p_F^a)^2\langle (\hat p_z)^2\rangle}{\left [\alpha_x+\beta_{xz}M_z^2+2\gamma^x (p_F^a)^2\right ]^2}- \frac{\gamma_{zz}^{z}(p_F^a)^2\langle (\hat p_z)^2\rangle}{\left [\alpha_y+\beta_{yz}M_z^2+2\gamma^y (p_F^a)^2\right ]^2} \right ).
\label{G2}
\end{eqnarray}
Unlike the paper \cite{Mineev2017} these expressions contain in denominator the factors $1+\lambda_a$  originating from the effective mass renormalisation. On the other hand in comparison with  \cite{Mineev2017} the Eqs. (\ref{G1}), (\ref{G2})  are simplified:
we  work with quadratic electron spectrum and spherical Fermi surface. Thus,
$N^a_0= m p_F^a/2\pi^2$ is the density of electronic states of the band $a=\uparrow,\downarrow$ without mass renormalisation due to spin-fluctuations, the average over Fermi surface 
$\langle (\hat p_z)^2\rangle=1/3$.
The function 
\begin{equation}
S(T)=2\pi T\sum_{n\geq 0}\frac{1}{\omega_n}=\ln\frac{\epsilon}{T},
\end{equation} 
$\epsilon=\frac{2\gamma\varepsilon_0}{\pi}$, $\ln\gamma=0.577$ is the Euler constant, and $\varepsilon_0$  is an energy cutoff for pairing interaction. We assume here that it has the same value for both bands.

 The zero of determinant of the system (\ref{Ax}) yields the BCS-type formula
\begin{equation}
T=\epsilon~\exp\left (-\frac{1}{\cal G}  \right ),
\label{21}
\end{equation}
where 
\begin{equation}
{\cal G}=\frac{{\cal G}_{1}^\uparrow+{\cal G}_{1}^\downarrow}{2}+\sqrt{\frac{({\cal G}_{1}^\uparrow-{\cal G}_{1}^\downarrow)^2}{4}+{\cal G}_{2}^{\uparrow}{\cal G}_{2}^{\downarrow}}
\end{equation}
is the function of temperature and  magnetic field. The Eq. (\ref{21}) is, in fact, an equation for the determination of the critical temperature of the transition to the superconducting state. 
At temperatures well below the Curie temperature one can neglect the temperature dependence of the coupling constant and Eq.(\ref{21})  determines the critical temperature of transition to superconducting state.

In the case of  a single-band (say spin-up) superconducting state, when ${\cal G}={\cal G}_{1}^\uparrow$, it is
\begin{equation}
T_{sc}=\epsilon~\exp\left (-\frac{1+\lambda_\uparrow}{N_0^\uparrow g^2\langle\chi_{zz}^{u\uparrow}(p_z,p_z)\rangle}  \right ).
\label{22}
\end{equation}
This formula reminds the known McMillan \cite{McMillan1968} expression $T_{sc}=\epsilon\exp\left (-\frac{1+\lambda}{\lambda}\right)$  valid for $s$-wave pairing in neglect  Coulomb repulsion. The expression similar to $s$-wave case was also obtained for the  transition temperature to $p$-wave superconducting state  in isotropic ferromagnet \cite{Fay1980}.
In our model the coefficient $\lambda_\uparrow$
determining the effective mass renormalisation does not coincide with the constant of interaction $N_0^\uparrow g^2\langle\chi_{zz}^{u\uparrow}\rangle$. The latter was derived in \cite{Mineev2017} taking into account
the orthorhombic anisotropy.

We have seen that   $\lambda_\uparrow$ decreases with magnetic field. $N^\uparrow_0= m p_F^\uparrow/2\pi^2$ increases
with magnetic field.
According to Eq.(\ref{G1})  the numerator in the formula for $\langle\chi_{zz}^{u\uparrow}\rangle$  is $\propto (p_F^\uparrow)^2$, hence, it increases with magnetic field whereas the denominator
 increases with field dependent  magnetisation. So, the critical temperature occurs magnetic field dependent quantity. 
  It should be borne in mind, however, that this dependence is weakened by the Coulomb repulsion which we neglected in our derivation.
  The temperature dependence of the upper critical field in UCoGe  in direction parallel to spontaneous magnetisation exhibits the peculiar upward curvature. 
   A natural explanation of this phenomenon proposed in \cite{Mineev2017}  is that the critical temperature itself is a decreasing function of the magnetic field.

\section{Upper critical field}

The upper critical field problem  for the B-state presents a two-band generalisation  of the corresponding problem for the superconducting polar state \cite{Book}.
The system of linear integral equations for determination of the upper critical field  along the spontaneous magnetisation ${\bf h}=h\hat z$ is
\begin{eqnarray}
\eta^\uparrow(\mbox{\boldmath$\rho$})=\hat L_1^\uparrow\eta^\uparrow(\mbox{\boldmath$\rho$})+\hat L_2^\downarrow\eta^\downarrow(\mbox{\boldmath$\rho$})
\nonumber\\
\eta^\downarrow(\mbox{\boldmath$\rho$})=\hat L_2^\uparrow\eta^\uparrow(\mbox{\boldmath$\rho$})+\hat L_1^\downarrow\eta^\downarrow(\mbox{\boldmath$\rho$}).
\label{C}
\end{eqnarray}
Here, the operators
\begin{eqnarray}
\hat  L_i^a f(\mbox{\boldmath$\rho$})=\frac{{\cal G}_i^a}{N_0^a\langle (\hat p_z)^2\rangle} \left ( \frac{m}{2\pi}\right)^2
T\sum_n \int_0^\infty dR
\exp \left(-\frac{2|\omega _n|}{v ^a_{F} }R \right)\nonumber\\
\times\int_0^\pi
 \sin \theta \cos ^2\theta d\theta \int_0^{2\pi}d\varphi \exp \left( i\frac{R}{{\sqrt 2}}
\sin \theta \left( e^{-i\varphi }D_+ +e^{i\varphi }D_-  \right) \right)
f(\mbox{\boldmath$\rho$}),
\end{eqnarray}
where $D_\pm=\frac{1}{\sqrt 2}(D_x\pm iD_y)$, ${\bf D}=-i\nabla+(2e/c){\bf A}({\bf r})$,  $curl{\bf A}={\bf B}={\bf h}+{\bf H}$, $\mbox{\boldmath$\rho$}=\rho(\cos\varphi,\sin\varphi)$.
By means the standard procedure \cite{Book} the system of integral equations is transformed  to the system of algebraic equations
\begin{eqnarray}
\eta^\uparrow={\cal G}^\uparrow_1I^\uparrow\eta^\uparrow+{\cal G}^\downarrow_2I^\downarrow\eta^\downarrow
\nonumber\\
\eta^\downarrow={\cal G}^\uparrow_2I^\uparrow\eta^\uparrow+{\cal G}^\downarrow_1I^\downarrow\eta^\downarrow,
\label{D}
\end{eqnarray}
where the integrals
\begin{eqnarray}
I^a =\frac{1}{N_0^a\langle (\hat p_z)^2\rangle} \left ( \frac{m}{2\pi}\right)^2
T\sum_n \int_0^\infty dR
\exp \left(-\frac{2|\omega _n|}{v ^a_{F} }R \right)
\int_0^\pi
 \sin \theta \cos^2\theta d\theta \exp\left ( -\frac{e (h+H)}{c}R^2\sin^2\theta\right)
 \end{eqnarray}
have the following property
\begin{equation}
I^\uparrow(h+H=0)=S(T),~~~~~~I^\downarrow(h+H=0)=S(T).
\end{equation}

Hence, the equations (\ref{D}) can be rewritten in the form eliminating of the logarithm divergency in the integrals $I^a$
\begin{eqnarray}
\eta^\uparrow={\cal G}^\uparrow_1\left [K^\uparrow+\frac{1}{\cal G}+\ln\frac{T_{sc}}{T}\right ]\eta^\uparrow+{\cal G}^\downarrow_2\left [K^\downarrow+\frac{1}{\cal G}+\ln\frac{T_{sc}}{T}\right ]\eta^\downarrow
\nonumber\\
\eta^\downarrow={\cal G}^\uparrow_2\left [K^\uparrow+\frac{1}{\cal G}+\ln\frac{T_{sc}}{T}\right ]\eta^\uparrow+{\cal G}^\downarrow_1\left [K^\downarrow+\frac{1}{\cal G}+\ln\frac{T_{sc}}{T}\right ]\eta^\downarrow,
\label{E}
\end{eqnarray}
where 
\begin{equation}
K^a=I^a-2\pi T\sum_{n\geq 0}\frac{1}{\omega_n}.
\end{equation}
Equating the determinant of this system to zero we obtain the equation for the determination of the upper critical field
\begin{equation}
\left \{{\cal G}^\uparrow_1\left [K^\uparrow+\frac{1}{\cal G}+\ln\frac{T_{sc}}{T}\right ]-1\right\}
\left \{{\cal G}^\downarrow_1\left [K^\downarrow+\frac{1}{\cal G}+\ln\frac{T_{sc}}{T}\right ]-1\right\}-
{\cal G}^\uparrow_2{\cal G}^\downarrow_2\left [K^\uparrow+\frac{1}{\cal G}+\ln\frac{T_{sc}}{T}\right ]\left [K^\downarrow+\frac{1}{\cal G}+\ln\frac{T_{sc}}{T}\right ]=0.
\label{33}
\end{equation}
 The solution of this equation at arbitrary temperature  taking into account the field dependence of all the quantities $G_i^a(H)$ and $T_{sc}(H)$ can be found  only numerically. Here, we will  obtain the analytic expressions  at $T=0$ and near the critical temperature.
 
The condition $h+H=0$ means that the external field in single-domain specimen completely compensates the internal field. At $h+H=B>0$  the temperature dependence of  $B_{c2}(T)$
corresponds to the temperature dependence of the upper critical field in poly-domain specimen at an external field exceeding the internal magnetisation \cite{Mineev2018}.

At $T\to 0$,  according to Ref.27,
\begin{equation}
K^a\approx\ln\left\{\frac{T}{T_{sc}} \left(\frac{e^{8/3}\phi_0}{8\pi\gamma(\xi_0^a)^2 
B_{c2}}\right)^{1/2}\right\},
\end{equation}
where
\begin{equation}
\xi_0^{a}=\frac{\hbar{v_F^a}}{2\pi T_{sc}}
\end{equation}
is the coherence length 
and $\phi_0=\frac{\pi\hbar c}{e}$ is the flux quantum.
Hence, in the Eq.(\ref{33})  the divergent  term  $\ln\frac{T_{sc}}{T}$ drops out and we obtain the quadratic equation in respect of $\ln\sqrt{B_{c2}}$. The solution of it  looks quite cumbersome. We write  here the corresponding expression for the one band  (spin-up) ferromagnet which coincides with the upper critical field for the polar phase \cite{Book}
\begin{equation}
B_{c2}(T=0)=\frac{e^{8/3}}{4\gamma}\frac{\phi_0}{2\pi(\xi_0^\uparrow)^2}.
\end{equation}
The coherence length $\xi_0^\uparrow$ itself is the magnetic field function.

Near the critical temperature when the $H+h=B \ll h$ 
\begin{equation}
K^a\approx-D^aB_{c2}=7\pi\zeta(3)\frac{(\xi_0^{a})^2}{\phi_0}B_{c2}.
\end{equation}
Hence, we obtain in linear in $\frac{T_{sc}-T}{T_{sc}}$ approximation
\begin{equation}
B_{c2}=\frac {({\cal G}^\uparrow_1-{\cal G}){\cal G}_1^\downarrow+({\cal G}^\downarrow_1-{\cal G}){\cal G}_1^\uparrow-2{\cal G}^\uparrow_2{\cal G}^\downarrow_2}
{({\cal G}^\uparrow_1-{\cal G}){\cal G}_1^\downarrow D^\downarrow+({\cal G}^\downarrow_1-{\cal G}){\cal G}_1^\uparrow D^\uparrow-{\cal G}^\uparrow_2{\cal G}^\downarrow_2
(D^\uparrow+D^\downarrow)}
\frac{T_{sc}-T}{T_{sc}}.
\end{equation}
For single band (say spin-up) ferromagnet the expression for the upper critical field acquires much simpler form
\begin{equation}
B_{c2}=\frac{\phi_0}{7\pi\zeta(3)(\xi_0^{\uparrow})^2}\frac{T_{sc}-T}{T_{sc}}.
\end{equation}
Here, $\zeta(x)$ is the Riemann zeta function.
The critical temperature $T_{sc}$ and the Fermi velocity $v_F^\uparrow$ are magnetic field dependent quantities. Hence, this expression presents an equation for the upper critical field determination.
As it was shown in \cite{Mineev2017}
when the critical temperature given by Eq.(\ref{22}) decreases with magnetic field  the temperature dependence of the upper critical field  acquires upward curvature. Here we see, that this effect 
is in fact even stronger due to increasing with field  the Fermi velocity $v_F^\uparrow$. The corresponding experimental plot is presented in Fig.1.

\begin{figure}[p]
\includegraphics
[height=1.0\textheight]
{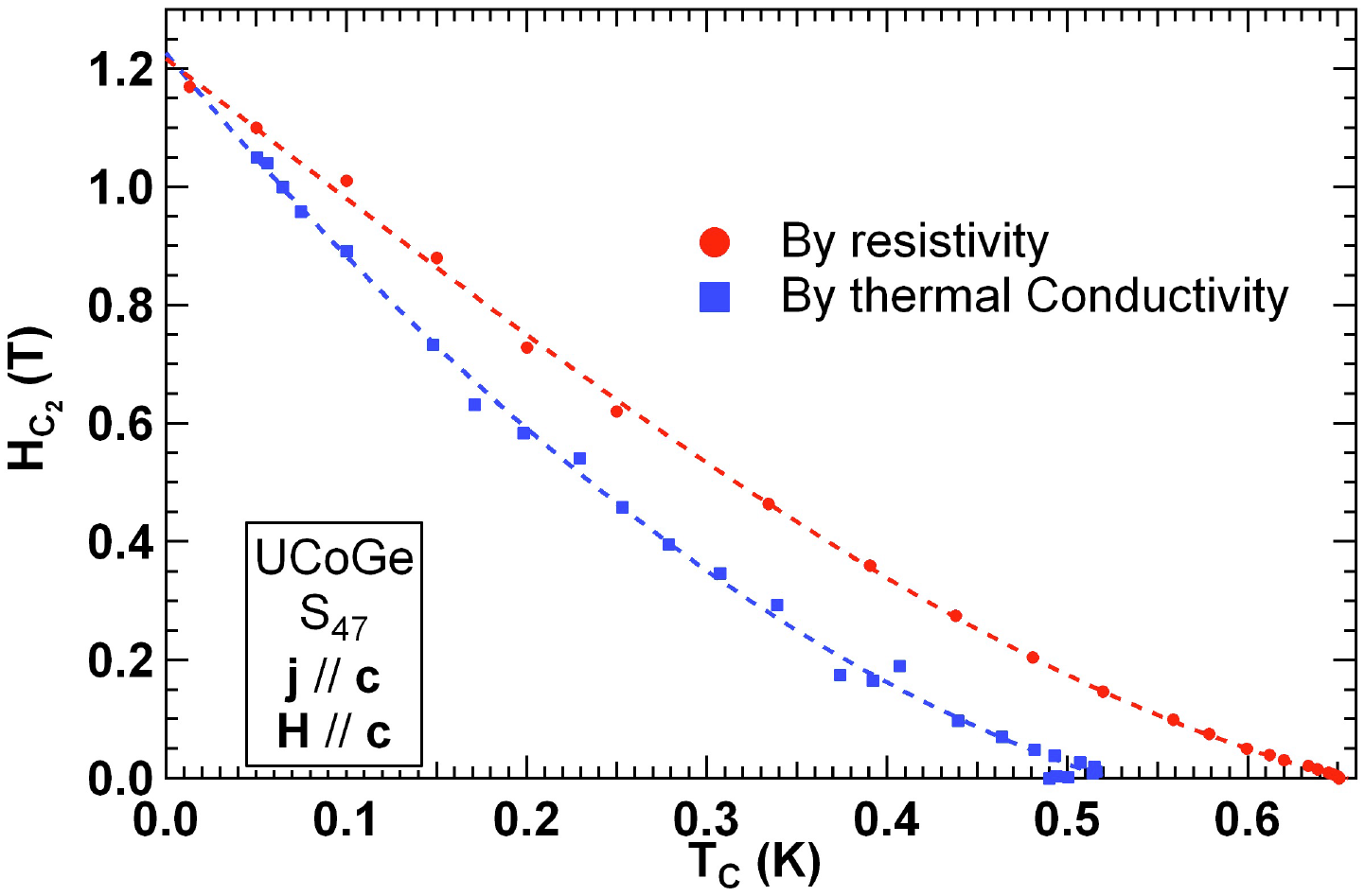}
 \caption{(Color online) 
The upper critical field 
$H_{c2}$ in UCoGe extracted from the resistivity and the thermal conductivity measurements.
(M.Taupin, unpublished (2016)). }
\end{figure}

\section{Conclusion}

We have studied the magnetic field dependence of effective mass and derived the equations for the temperature dependence of the upper critical field 
in a two band ferromagnetic superconductor with triplet pairing. It was found that
the low temperature specific heat is slowly decreasing  function of magnetic field along the easy axis.
 This corresponds to the experimental observations \cite{Aoki2019}. 
There was shown that the  behaviour of the upper critical field is not the same as in usual two band superconductor. The fact is that the pairing coupling itself and the bands Fermi velocities are not a constants but proved
to be magnetic field dependent quantities.
The field dependence of the interaction constant is determined by the field dependence of Fermi surface radii and the magnetisation  increasing in   magnetic field directed along the spontaneous magnetisation. 
Unlike to s-wave superconductivity the mass renormalisation coefficient $\lambda$ does not coincide with the pairing interaction constant derived taking into account the orthorhombic anisotropy. Due dependencies of the  critical temperature and the Fermi velocity from the magnetic field   the upper critical field can exhibit the upward curvature.

Finally I would like to mention the principal simplifications made in the calculations.
The derivation have been done for two band spin-up spin-down ferromagnetic superconductor with equal-spin pairing. 
The effective mass and the pairing coupling constants were derived by the field theoretical method but  making use the phenomenological formulae for the spin susceptibility components.
I have worked with isotropic electron and spin-fluctuation spectra. The latter assumptions allowing  to perform calculation analytically are obviously inconsistent with orthorhombic structure of ferromagnetic compounds under consideration. The Coulomb electron-electron interaction has been neglected. 
Despite of the shortcomings the presented theory allows to establish the qualitative field dependence of measurable quantities such as the electron effective mass and the upper critical field in ferromagnetic superconductors.

\end{document}